\documentclass{svproc}
\usepackage{url,graphicx}

\usepackage{graphicx}
\begin{document}
\mainmatter              
\title{Neutron capture cross sections from surrogate reaction data and theory: connecting the pieces with a Markov-Chain Monte Carlo approach}
\titlerunning{MCMC Surrogate Method}  
%
\author{Oliver Gorton \inst{1} \and Jutta E. Escher \inst{2}}
\authorrunning{Oliver Gorton et al.} 
%
\tocauthor{Oliver Gorton, Jutta Escher}
\institute{San Diego State University, San Diego, CA 92108, USA,\\
\email{ogorton@sdsu.edu},\\ \and
Lawrence Livermore National Laboratory,
7000 East Avenue,\\
Livermore, CA 94550, USA\\
\email{escher1@llnl.gov}}

\maketitle              

\begin{abstract}
The neutron capture cross section for $^{90}Zr(n,\gamma)$ has recently been determined
using surrogate $^{92}Zr(p,d\gamma)$ data and nuclear reaction theory \cite{prl}. That work 
employed an approximate fitting method based on Bayesian Monte Carlo sampling to determine
parameters needed for calculating the $^{90}Zr(n,\gamma)$ cross section. Here, we improve the
approach by introducing a more sophisticated Markov Chain Monte Carlo sampling method~\cite{hast}.
We present preliminary results.
\end{abstract}

Neutron capture cross sections can be measured by bombarding 
a sample of target nuclei with neutrons and detecting decay products. 
Such measurements cannot be completed in the laboratory when 
the target isotopes have half-lives that are short compared to 
timescales relevant to the experiment. This leaves critical gaps in 
nuclear data libraries. To predict the missing data, nuclear cross 
section calculations can, in principle, be carried out 
using statistical Hauser-Feshbach (HF) models \cite{capote}. 
In compound nuclear reactions, a compound nucleus (CN) is formed, which then decays
through the available decay channels. These channels and the probability of 
each being taken depends on
the nuclear level densities and $\gamma$-ray strength function of the CN.
The general lack of nuclear structure information for medium to heavy mass nuclei
leads to the need for indirect constraints on the corresponding HF parameters. 
The surrogate method \cite{esch} obtains these constraints using measurements 
of the same CN decay observed in alternative reactions.

Specifically, in Ref.~\cite{prl} the decay of the CN $^{91}Zr$ was modeled using parame\-trized
(phenomenological) forms for the level density and $\gamma$-ray strength function. 
The parameters were fitted to measured $^{92}Zr(p,d\gamma)$ data from a surrogate experiment
and subsequently used to calculate the desired $^{90}Zr(n,\gamma)$ cross section.
A Bayesian Monte Carlo approach was employed, which provided an average $(n,\gamma)$
cross section, along with a variance, yielding an uncertainty band that is symmetric around the
mean.
Here, we improve the parameter estimation by introducing a Markov-Chain Monte Carlo (MCMC) approach for 
sampling the HF parameter space, generating a joint probability distribution for the parameters 
without visiting every combination of parameters. 

The nuclear level density model we employ is the composite Gilbert-Cameron level density \cite{gcld} 
with the Ignatyuk treatment of the energy dependence of the level density parameter \cite{igna}. We varied five parameters within this
presription, which are, following the notation of reference \cite{capote}: the asymptotic level density parameter
$\tilde{a}$, the shell correction energy $\delta W$, the level density damping parameter $\gamma$, the pairing
energy shift $\Delta$, 
and the effective moment of inertia that enters the expression for the spin-cutoff factor.
The $\gamma$-ray strength function description employed was the enhanced generalized Lorentzian (EGLO)
model for the E1 transitions, and the standard Lorentzian (SLO) for the M1 transitions.
These models are parameterized by their peak energy, width, and strength \cite{capote}.
We varied nine strength function parameters, three for each peak, with the EGLO function having two peaks, and the
SLO having a single peak. A total of 14 parameters were varied simultaneously.

\begin{figure}
	\centering
	\includegraphics[width=\textwidth]{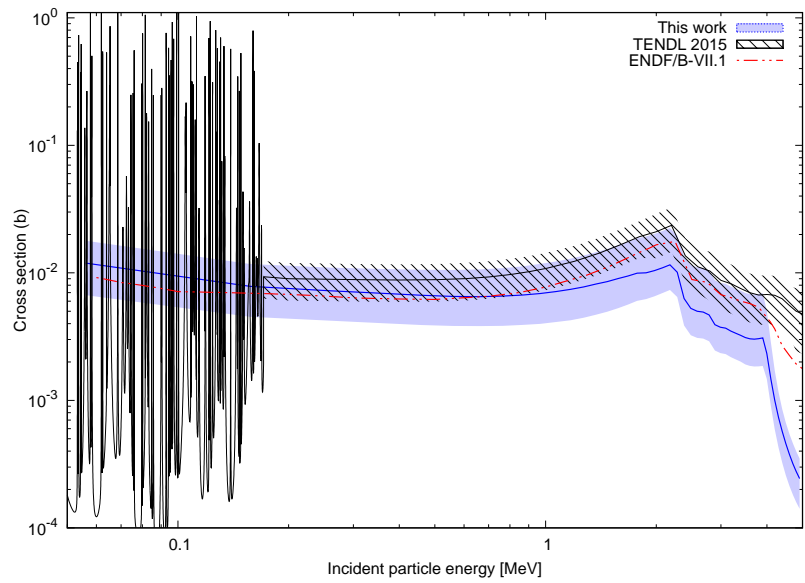}
	\caption{Preliminary $^{90}Zr(n,\gamma)$ cross section obtained indirectly from 
	$^{92}Zr(p,d\gamma)$ data using the newly-developed MCMC approach. The solid (blue) curve is
	the median value and the solid (blue) band indicate the 68\% confidence interval. 
	For comparison, the Talys Evaluated Nuclear Data Library (TENDL) \cite{koning} and the Evaluated
	Nuclear Data File (ENDF) library results are shown a well \cite{endf}.}
	\label{nccs}
\end{figure}

We employ a Metropolis-Hastings MCMC algorithm \cite{hast,mcmc}, and explore convergence of the sampling process.
The prior distributions for each parameter was finite and flat, and centered around recommended parameter values from RIPL-3 \cite{capote}.
The posterior parameter distribution we obtain is sampled, yielding the $^{90}Zr(n,\gamma)$ cross section shown 
in Figure~\ref{nccs}.
This method propagates all constraints encoded in the joint probability distribution of parameters, including correlations between the parameters.
The result is seen to be in agreement with the TENDL 2015 and ENDF/B-VII.1 evaluations, both of 
which are based on directly-measured data. 
In future work we will investigate correlations between the model parameters and 
between the cross sections at different energies.

{\bf Acknowledgments.} This work was performed under the auspices of the U.S. Department of Energy by Lawrence Livermore National Laboratory under contract DE-AC52-07NA27344, with support from LLNL's HEDP summer student program and LDRD project 19-ERD-017.

\end{document}